\documentclass[a4paper,10pt,aps,pre,twocolumn,superscriptaddress,showpacs,amsmath,amssymb]{revtex4}
\usepackage{graphicx,color}

\newcommand{\dd}{\mathrm{d}}
\newcommand{\td}[2]{\frac{\dd #1}{\dd #2}}
\newcommand{\pd}[2]{\frac{\partial #1}{\partial #2}}
\newcommand{\fd}[2]{\frac{\delta #1}{\delta #2}}
\newcommand{\mean}[1]{\langle #1 \rangle}
\newcommand{\Int}[1]{\int\dd #1\;}
\newcommand{\IInt}[3]{\int_{#2}^{#3}\dd #1\;}
\newcommand{\PInt}[1]{\int[\dd #1]\;}

\renewcommand{\vec}[1]{\mathbf #1}

\DeclareMathOperator{\erfc}{erfc}

\newcommand{\Gam}{\Gamma}
\newcommand{\eps}{\varepsilon}
\newcommand{\kap}{\kappa}
\newcommand{\lam}{\lambda}

\newcommand{\im}{\text i}
\newcommand{\ra}{\rightarrow}
\newcommand{\x}{\vec r}
\newcommand{\X}{\vec R}
\newcommand{\nois}{\boldsymbol\xi}

\newcommand{\Fid}{F_\text{id}}
\newcommand{\Ns}{N_\text{s}}
\newcommand{\tr}{\tau_\text{r}}
\newcommand{\pc}{p_\text{co}}
\newcommand{\pe}{p_\text{ex}}

\begin{document}

\title{Stochastic thermodynamics of fluctuating density fields: \\
  Non-equilibrium free energy differences under coarse-graining}

\author{T. Leonard}
\author{B. Lander}
\author{U. Seifert}
\affiliation{II. Institut f\"ur Theoretische Physik, Universit\"at Stuttgart,
  Pfaffenwaldring 57, 70550 Stuttgart, Germany}
\author{T. Speck}
\affiliation{Institut f\"ur Theoretische Physik II,
  Heinrich-Heine-Universit\"at D\"usseldorf, Universit\"atsstra\ss e 1, 40225
  D\"usseldorf, Germany}

\begin{abstract}
  We discuss the stochastic thermodynamics of systems that are described by a
  time-dependent density field, for example simple liquids and colloidal
  suspensions. For a time-dependent change of external parameters, we show
  that the Jarzynski relation connecting work with the change of free energy
  holds if the time evolution of the density follows the Kawasaki-Dean
  equation. Specifically, we study the work distributions for the compression
  and expansion of a two-dimensional colloidal model suspension implementing a
  practical coarse-graining scheme of the microscopic particle positions. We
  demonstrate that even if coarse-grained dynamics and density functional do
  not match, the fluctuation relations for the work still hold albeit for a
  different, apparent, change of free energy.
\end{abstract}


\maketitle


\section{Introduction}

The estimation of absolute and relative free energies lies at the very heart
of computational chemistry. Accordingly, a wide range of methods such as
thermodynamic integration and umbrella sampling has been developed, see, e.g.,
Refs.~\citenum{ytre06,poho10,chod11} for recent reviews. With the advent of
fluctuation theorems~\cite{gall95,kurc98,lebo99,seif05a} and the Jarzynski
relation~\cite{jarz97}, non-equilibrium methods have been added to this
arsenal. These non-equilibrium methods promise to extract free energy
differences from switching trajectories that are harvested out of
equilibrium. It has been argued that non-equilibrium methods generally are
less efficient than conventional methods~\cite{ober05}. There might be,
however, situations where they constitute the only possibility, e.g. in
force-probe experiments and molecular dynamics simulations of large
biomolecules~\cite{merk99}. Moreover, the free energy estimated from
trajectories generated under uni-directional protocols is biased, a problem
that can be alleviated by bi-directional sampling schemes employing Bennett's
acceptance ratio method~\cite{benn76,shir03,kim12}.

An open question in stochastic thermodynamics~\cite{seif12} is how
coarse-graining and hidden degrees of freedom influence the statistics, and
therefore the fluctuation theorems, of work and
heat~\cite{raha07,pugl10,alta12,alta12a,espo12,mehl12,cela12,kawa13}. For
example, overdamped dynamics is typically used to model soft matter systems on
time scales larger than the relaxation time of momenta. In this case a reduced
description is found by eliminating fast degrees of freedom. However, in an
environment with a spatially varying temperature there is a contribution to
the entropy production (due to microscopic cycles) that is missed by the
overdamped dynamics~\cite{cela12}. Another possibility is to eliminate slow
degrees of freedom: In Ref.~\citenum{mehl12} the position of a colloidal
particles has been measured that is coupled to another ``hidden'' particle,
leading to a modified slope of the fluctuation theorem. For discrete state
spaces, coarse-graining based on a time-scale separation has been discussed
for removing fast states~\cite{pugl10,alta12,alta12a} and the clustering of
microstates into mesostates~\cite{raha07,espo12}.

In this paper, we consider another kind of coarse-graining arising in
many-body systems, where we might no longer be able (or interested) to fully
resolve the microscopic particle positions. Instead, we measure a
coarse-grained density profile by counting particles in a probe volume over
some time. This density fluctuates, i.e., repeating the experiment we will
obtain slightly different densities for each realization. Such a behavior
calls for a field theoretical treatment. In Sec.~\ref{sec:st}, we make the
link to stochastic thermodynamics by proving the Jarzynski relation under two
conditions: (i) The time evolution of density fluctuations are governed by the
Kawasaki-Dean equation~\cite{kawa93,dean96} and (ii) the free energy
functional that enters the Kawasaki-Dean equation is given by the same
functional that determines the equilibrium weight of density fluctuations away
from the most probable density profile. There is a close link to density
functional theory~\cite{evan79,lowe94,roth10}, which has been applied
successfully to study the structural and thermodynamics properties of a wide
range of spatially inhomogeneous systems. As an illustration, in
Sec.~\ref{sec:illu} we consider the compression and expansion of a colloidal
suspension. While this driving protocol has been studied
previously~\cite{lua05,croo07,davi12,hopp13}, here we employ it to investigate
the effects of coarse-graining on the work distribution and the resulting
change of free energy.


\section{Stochastic thermodynamics of coarse-grained densities}
\label{sec:st}

\subsection{Coarse-grained densities}

We consider a system of particles confined in a volume $V$ in contact with a
heat reservoir at temperature $T$ (throughout we set Boltzmann's constant to
unity). For simplicity, we restrict ourselves to a constant number of
particles $N$ but the extension to a particle reservoir is straightforward. We
assume that the density $\rho(\x)$ is the result of some coarse-graining
procedure in space, in time, or both, of the microscopic degrees of
freedom. Hence, $\rho(\x)$ fluctuates, and fluctuations away from the most
probable profile are governed by the Boltzmann distribution
\begin{equation}
  \label{eq:boltz}
  \psi[\rho] = e^{-(F[\rho]-\Phi)/T}
\end{equation}
with free energy functional $F[\rho]$. This functional follows from the
definition of the Helmholtz free energy $\Phi$ through
\begin{equation}
  \label{eq:helm}
  e^{-\Phi/T} \equiv \Int{\x_1\cdots\dd\x_N} e^{-\hat U/T} = \PInt{\rho} e^{-F[\rho]/T}
\end{equation}
as
\begin{equation}
  F[\rho] = -T\ln\Int{\x_1\cdots\dd\x_N} \delta[\rho(\x)-\rho_\text{CG}(\x)]
  e^{-\hat U/T},
\end{equation}
where $\hat U(\{\x_k\})$ is the potential energy and
$\rho_\text{CG}(\x|\{\x_k\})$ formalizes the coarse-graining procedure through
mapping a set of coordinates onto a density field. The free energy functional
$F[\rho]=\Fid[\rho]+U[\rho]$ of a given density profile $\rho(\x)$ is
customarily split into an ideal gas part
\begin{equation}
  \label{eq:id}
  \Fid[\rho] \equiv T\Int{\x}\rho(\x)[\ln\rho(\x)-1]
\end{equation}
and the excess free energy $U[\rho]$, which is not known explicitly in
general.

In order to proceed, we need to provide an equation of motion that governs the
temporal evolution of the density field $\rho(\x,t)$. This has been a point of
some debate as reviewed by Archer and Rauscher~\cite{arch04}. Let us first
consider the \emph{microscopic} density
\begin{equation}
  \label{eq:rho:m}
  \hat\rho(\x,t) \equiv \sum_{k=1}^N \delta(\x-\x_k(t)),
\end{equation}
where $\x_k$ denotes the position of the $k$th particle. Here, no
coarse-graining has been performed and we assume that we have full knowledge
of all particle positions. In this case the excess part of the density
functional equals the potential energy, $U[\hat\rho]=\hat
U=\sum_{k<l}u(|\x_k-\x_l|)$ (assuming pairwise interactions), and can thus be
expressed through the quadratic form
\begin{equation}
  \label{eq:U}
  U[\rho] \equiv \frac{1}{2}\Int{\x\dd\x'} \rho(\x)u(|\x-\x'|)\rho(\x').
\end{equation}
Starting from the overdamped stochastic motion of $N$ particles interacting
via the pair potential $u(r)$, Dean~\cite{dean96} has shown that the
microscopic density Eq.~\eqref{eq:rho:m} exactly obeys the equation of motion
\begin{equation}
  \label{eq:dean}
  \partial_t\hat\rho = 
  \nabla\cdot\left[\hat\rho\nabla\fd{F[\hat\rho]}{\hat\rho} + \nois \right],
\end{equation}
where the noise has zero mean and correlations
\begin{equation}
  \label{eq:corr}
  \mean{\xi_i(\x,t)\xi_j(\x',t')} =
  2T\hat\rho(\x,t)\delta_{ij}\delta(\x-\x')\delta(t-t').
\end{equation}
The noise is thus multiplicative and vanishes in regions where the density is
zero, i.e., particles are absent.

It is instructive to emphasize the difference to classical density functional
theory (DFT)~\cite{regu04}, which is based on the functional
\begin{equation}
  \mathcal F[\rho] \equiv \min_{\Psi|\rho} \Int{\x_1\cdots\dd\x_N}
  \Psi[\hat U+T\ln\Psi],
\end{equation}
where $\Psi(\{\x_k\})$ is the full many-body distribution. This functional is
obtained as a constrained minimization over all normalized distributions
$\Psi$ that yield the density profile
$\rho(\x)=\Int{\x_1\cdots\dd\x_N}\Psi\hat\rho$. In particular, the equilibrium
density $\mean{\hat\rho}_0$ fulfills
\begin{equation}
  \left.\fd{\mathcal F[\rho]}{\rho}\right|_{\mean{\hat\rho}_0} = \mu,
\end{equation}
where $\mu$ is the chemical potential. The free energy functional $\mathcal
F[\rho]$ can again be split into the ideal part Eq.~\eqref{eq:id} and an
excess part. The latter is not known in general, however, excellent
approximations have been obtained for, e.g., hard spheres (see
Ref.~\citenum{roth10} and references therein). Since densities in DFT are
ensemble averages, dynamical DFT results in a deterministic equation
\begin{equation}
  \label{eq:ddft}
  \partial_t\rho = \nabla\cdot\left[\rho\nabla\fd{\mathcal
      F[\rho]}{\rho}\right]
\end{equation}
lacking the noise term from Eq.~\eqref{eq:dean}. Note that Eq.~\eqref{eq:ddft}
follows from an adiabatic approximation assuming that the two-body density in
the time-dependent case is that of the stationary equilibrium system at the
same one-body density $\rho(\x)$. Non-adiabatic corrections have been
discussed recently within a variational approach based on Rayleigh's
dissipation functional~\cite{schm13a}.

Here, we are interested in the case of fluctuating densities $\rho(\x,t)$ that
are the result of a coarse-graining procedure. We assume that
Eq.~\eqref{eq:dean} in conjunction with the correct free energy functional,
which we will refer to as the Kawasaki-Dean equation, holds for the temporal
evolution of such a coarse-grained density profile. For the microscopic
density $\hat\rho$ using Eq.~\eqref{eq:U}, this is an exact result. Note that
the free energy functional $F[\rho]$, in general, is different from $\mathcal
F[\rho]$ and does depend on the details of the coarse-graining procedure,
i.e., the excess part will be different from the quadratic form
Eq.~\eqref{eq:U}.

\subsection{Stochastic thermodynamics}

Suppose that we drive the system out of thermal equilibrium by changing one or
more external parameters, which we denote $\lambda$. The total change of the
free energy functional
\begin{equation}
  \label{eq:fl}
  \td{F[\rho]}{t} = \Int{\x}\fd{F[\rho]}{\rho(\x,t)}\partial_t\rho(\x,t) +
  \pd{F[\rho]}{\lam}\dot\lam \equiv \dot q_0 + \dot w
\end{equation}
can be split into two terms, from which we identify the second term as the
work rate $\dot w$. The heat $\dot q=\dot q_0+T\td{}{t}S$ dissipated into the
reservoir stems from two sources: from the change of the free energy
functional and from the change of the constrained intrinsic entropy $S[\rho]$
associated with the set of different microstates corresponding to the density
profile $\rho$, see Ref.~\citenum{seif12} for a detailed discussion. We
introduce trajectories as histories of density profiles in time,
$\Gam\equiv\{\rho(\x,t):0\leqslant t\leqslant\tau\}$, and a protocol $\lam(t)$
that describes the switching between initial, $\lam(0)=\lam_0$, and final,
$\lam(\tau)=\lam_1$, state.

The central quantity that enters the derivation of fluctuation theorems is the
ratio of the probabilities $\mathcal P$ for forward and backward trajectories,
\begin{equation}
  \label{eq:R}
  \mathcal R[\Gam;\lam] \equiv \ln\frac{\mathcal P[\Gam;\lam]}{\mathcal
    P[\Gam^\dagger;\lam^\dagger]},
\end{equation}
where $\Gam^\dagger=\{\rho(\x,\tau-t):0\leqslant t\leqslant\tau\}$ and
$\lam^\dagger(t)=\lam(\tau-t)$ denote the time reversal of trajectory and
protocol, respectively. The weight $\mathcal P[\Gam;\lam]$ of a single
trajectory depends on the dynamics of the density fluctuations. Assuming that
the temporal evolution Eq.~\eqref{eq:dean} still holds for a coarse-grained
density $\rho(\x)$ with the appropriate free energy functional $F[\rho]$, we
show in appendix~\ref{sec:ft} that the ratio Eq.~\eqref{eq:R} becomes
\begin{equation}
  \label{eq:R:res}
  \mathcal R = -\frac{q_0}{T} 
  + \ln\frac{\psi_0[\rho(\x,0)]}{\psi_1[\rho(\x,\tau)]},
\end{equation}
where
\begin{equation}
  \label{eq:q}
  q_0 \equiv \IInt{t}{0}{\tau} \dot q_0 = 
  \Int{t\dd\x}\fd{F[\rho]}{\rho(\x,t)}\partial_t\rho(\x,t).
\end{equation}
The boundary term is given by the equilibrium Boltzmann distributions
$\psi_\lam$, Eq.~\eqref{eq:boltz}, of initial and final state.

Equipped with the ratio Eq.~\eqref{eq:R:res}, a number of results can be
obtained following the standard approach~\cite{seif12}. The arguably most
prominent is the Jarzynski relation~\cite{jarz97}
\begin{equation}
  \label{eq:jr}
  \mean{e^{-w/T}} = \PInt{\Gam} e^{-w/T}\mathcal P = e^{-\Delta\Phi/T},
\end{equation}
which follows through combining Eqs.~\eqref{eq:R}, \eqref{eq:R:res}, and
\eqref{eq:boltz} with $F_1-F_0=q_0+w$ [Eq.~\eqref{eq:fl}]. Eq.~\eqref{eq:jr}
relates the average over non-equilibrium trajectories on the left hand side
with the change $\Delta\Phi\equiv\Phi_1-\Phi_0$ of the equilibrium Helmholtz
free energy Eq.~\eqref{eq:helm} between final and initial state on the right
hand side.

The most common approach to calculate the free energy difference $\Delta\Phi$
between two states is thermodynamic integration~\cite{frenkel}, which
corresponds to a quasi-static process ($\dot\lam\ra 0$). In practice, one
performs many equilibrium simulations at slightly different values of $\lam$
and integrates the mean energy along this path from initial to final
$\lam$. Using Eq.~\eqref{eq:jr}, one could also calculate the free energy
difference from trajectories at finite switching speed $\dot\lam\neq0$. There
is a severe caveat one encounters trying to implement such a protocol: For a
finite number $\Ns$ of trajectories, the estimator of the average
\begin{equation}
  \mean{e^{-w/T}} \simeq \frac{1}{\Ns}\sum_{n=1}^{\Ns} e^{-w_n/T}
\end{equation}
is dominated by rare events having a large weight due to the exponential,
where $w_n$ is the work along the $n$th trajectory. Two ways out are to either
use approximations for the work distribution or to employ bi-directional
sampling. For finite but slow driving speeds $\dot\lam$ the work distribution
approaches a Gaussian~\cite{spec04,spec11a} but extreme tails, dominating the
average Eq.~\eqref{eq:jr}, might be
non-Gaussian~\cite{nick11,hopp13a}. Bi-directional sampling uses work values
obtained for both the forward and the reverse protocol. The free energy
difference is calculated through solving
\begin{equation}
  \label{eq:bennett}
  \Delta\Phi =
  -T\ln\frac{\sum_n[1+e^{(w_n-\Delta\Phi)/T}]^{-1}}
  {\sum_n[e^{w^\dagger_n/T}+e^{-\Delta\Phi/T}]^{-1}}
\end{equation}
iteratively. Here, $w_n$ is the work along the $n$th trajectory under the
forward protocol and $w_n^\dagger$ is the work along the $n$th trajectory
under the reverse protocol. For simplicity we have assumed an equal number of
trajectories for each protocol. In practice, for this procedure to converge,
the work distributions for the two protocols need to overlap. Hence, even
though in principle Eq.~\eqref{eq:jr} is valid for any driving protocol, in
practice both the approximated distribution and the bi-directional sampling
require a sufficiently slow driving speed.

\subsection{Compression and expansion}

We drive the suspension by compressing and expanding the occupied volume
$V=\lam^d$ at constant particle number $N$, where $d$ is the number of
dimensions. These two protocols connect two state points in the phase diagram
with equal temperatures. The control parameter $\lam$ now denotes the edge
length of the volume. To calculate the work rate, we rescale lengths
$\x\mapsto\lam\X$ with
\begin{multline}
  F[\rho] = T\Int{\X}\rho(\X)[\ln\rho(\X)/\lam^d - 1] \\
  + \frac{1}{2}\Int{\X\dd\X'} \rho(\X)u(\lam|\X-\X'|)\rho(\X')
\end{multline}
leading to
\begin{equation}
  \dot w[\rho] = \pd{F[\rho]}{\lam}\dot\lam = -P[\rho]\dot V.
\end{equation}
The incremental work thus has the expected form of a pressure times the volume
change. The pressure is a fluctuating observable since we control the volume.

In order to obtain a concrete expression for the pressure, we employ the
quadratic form Eq.~\eqref{eq:U} for the excess part of the density functional
$F[\rho]$. After we have restored the original lengths, the pressure reads
\begin{equation}
  \label{eq:P}
  P[\rho] \equiv \frac{TN}{V} + 
  \frac{1}{2dV}\Int{\x\dd\x'} \rho(\x,t)f(|\x-\x'|)\rho(\x',t)
\end{equation}
with $f(r)\equiv-ru'(r)$, where the prime denotes the derivative with respect
to the argument. The first term stems from the ideal gas part,
Eq.~\eqref{eq:id}, of the free energy. The second part incorporates the
interactions between particles. Inserting the microscopic density
Eq.~\eqref{eq:rho:m}, it reduces to the usual microscopic expression for the
virial.

\section{Colloidal suspension}
\label{sec:illu}


\subsection{Simulation details}

As a specific illustration, we consider a model suspension of $N=400$
colloidal particles moving in two dimensions. The particles interact via the
Yukawa potential
\begin{equation}
  \label{eq:yuk}
  u(r) = \eps\frac{e^{-\kap r}}{r}
\end{equation}
with screening length $\kap^{-1}$ and strength of the potential $\eps$, which
is set to $\eps=1$. Eq.~\eqref{eq:yuk} follows from DLVO theory~\cite{israel}
and is a common model for charge-stabilized colloidal particles. Throughout
this section, we report all energies in units of the thermal energy. Moreover,
as units of length and time we use $\kap^{-1}$ and $(D_0\kap^2)^{-1}$,
respectively, where $D_0$ is the short-time diffusion coefficient. We employ
Brownian dynamics simulations with periodic boundaries. Particle positions are
updated through integrating the coupled Langevin equations
\begin{equation}
  \label{eq:lang}
  \dot\x_k = -\sum_{l\neq k}u'(|\x_{kl}|)\frac{\x_{kl}}{|\x_{kl}|} + \nois_k
\end{equation}
with time step $\Delta t=10^{-3}$, where $\x_{kl}\equiv\x_k-\x_l$ and the
noise has correlations
$\mean{\xi_{ki}(t)\xi_{lj}(t')}=2\delta_{ij}\delta_{kl}\delta(t-t')$. For the
force evaluation, we use a cut-off radius $r_\text{c}=5$.

\subsection{Gaussian smearing}

For the coarse-grained density, we map the particle coordinates onto the
sum of Gaussians
\begin{equation}
  \label{eq:rho}
  \rho_\text{CG}(\x) \equiv \sum_{k=1}^N (2\pi\ell^2)^{-d/2}
  \exp\left\{-\frac{|\x-\x_k|^2}{2\ell^2}\right\},
\end{equation}
where $\ell$ is the coarse-graining length over which particle positions are
smeared out. In the limit $\ell\ra0$, we recover the microscopic density
$\hat\rho$, Eq.~\eqref{eq:rho:m}.
Inserting the explicit expression Eq.~\eqref{eq:rho} for the density profile
into Eq.~\eqref{eq:P}, we can simplify the coarse-grained pressure to a sum
over particle pairs. To this end, we employ the Fourier transform of the pair
potential $u(r)$ and write
\begin{equation}
  f(r) = -\frac{1}{(2\pi)^d}\Int{\vec q} u(q) \vec q\cdot\nabla_{\vec q}
  e^{\im\vec q\cdot\x}.
\end{equation}
We can now perform the integrals over the densities,
\begin{equation}
  \Int{\x}\rho(\x,t)e^{\im\vec q\cdot\x} = \sum_{k=1}^N\exp\left\{-\frac{1}{2}(\ell
    q)^2+\im\vec q\cdot\x_k\right\},
\end{equation}
which suggest to introduce the function $u_\ell(q)\equiv u(q)e^{-(\ell
  q)^2}$. Performing the differentiation with respect to $\vec q$ and putting
everything together, we find for the coarse-grained pressure
$P[\rho(\x,t)]=P_\ell(t)$ with
\begin{equation}
  \label{eq:P:ell}
  P_\ell = \frac{N}{V} + \frac{1}{dV}\sum_{k<l} f_\ell(|\x_k-\x_l|),
\end{equation}
where we have introduced the effective ``two-body pressure''
\begin{equation}
  \label{eq:f}
  f_\ell(r) \equiv -ru_\ell'(r) - 2\ell^2[u_\ell''(r)+u_\ell'(r)/r]
\end{equation}
and $u_\ell(r)$ is the back-transformation of $u_\ell(q)$. Clearly, for
$\ell=0$ we recover $f_0(r)=f(r)$.

\begin{figure}[t]
  \centering
  \includegraphics{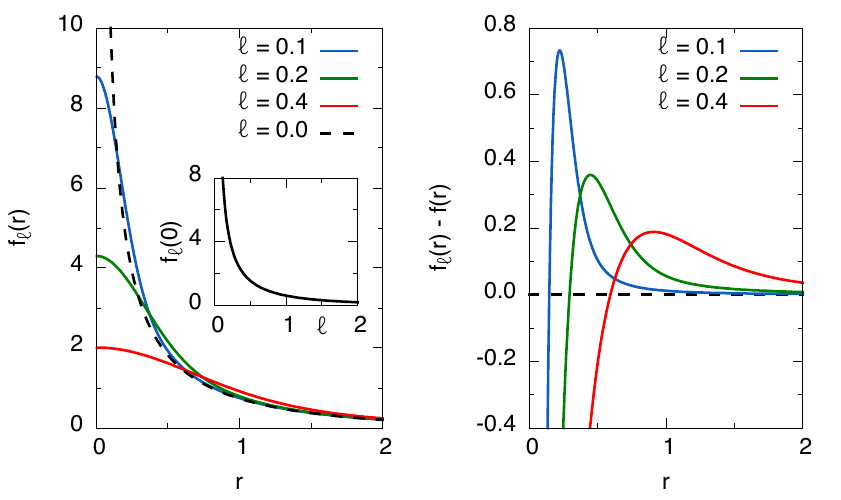}
  \caption{(Left)~The effective two-body pressure $f_\ell(r)$ from
    Eq.~\eqref{eq:f} as a function of distance $r$ for three coarse-graining
    lengths $\ell$ (solid lines) and the function $f(r)$ (dashed line). The
    inset shows the value at $r=0$ as a function of $\ell$. (Right)~The
    difference $f_\ell(r)-f(r)$ is negative for small distances $r$, passes
    through a maximum, and then approaches zero for large distances. For
    larger coarse-graining length $\ell$, the distance at which zero is
    crossed also becomes larger.}
  \label{fig:press}
\end{figure}

The advantage of Eq.~\eqref{eq:P:ell} is that we can calculate the
coarse-grained pressure (and therefore the work) for a range of
coarse-graining lengths from a single simulation integrating
Eq.~\eqref{eq:lang}. Inserting for $d=2$ dimensions the known Fourier
transform of Eq.~\eqref{eq:yuk}, we obtain
\begin{equation}
  \label{eq:u}
  u_\ell(r) = \IInt{q}{0}{\infty} \frac{q}{\sqrt{1+q^2}} e^{-(\ell q)^2} J_0(qr),
\end{equation}
where $J_0(x)$ is the zero-order Bessel function of the first kind. Taking the
derivative with respect to $r$ leads to integrals involving higher-order
Bessel functions. For a predefined set of lengths $\ell$, we numerically
evaluate the resulting integrals and tabulate the functions Eq.~\eqref{eq:f}
required for calculating the coarse-grained pressure from a given particle
configuration. In Fig.~\ref{fig:press}, we plot $f_\ell(r)$ and the difference
$f_\ell(r)-f(r)$. While $f(r)$ diverges for $r\ra0$, this divergence is
removed for any $\ell>0$ and the apparent pressure $f_\ell(0)$ between
overlapping particles remains finite. The inset shows this value
\begin{equation}
  f_\ell(0) = 1 - \sqrt\pi\frac{2\ell^2-1}{2\ell} e^{\ell^2}\erfc(\ell)
\end{equation}
as a function of $\ell$, where $\erfc(x)$ denotes the complementary error
function. The functions $f_\ell(r)$ start from a finite value but cross $f(r)$
and then approach $f(r)$ from above. Hence, there is an intermediate range in
which the effective two-body pressure is increased.


\subsection{Work distributions}

We employ bi-directional sampling. We simulate $\Ns$ cycles during which we
record one work value for the compression and one for the expansion process:
During the time $\tr$ the system is allowed to relax at the initial number
density $\rho_0=N/\lam_0^2$ and fixed $\lam_0$. The system is then compressed
to a higher density $\rho_1=N/\lam_1^2$ during the switching time $\tau$. To
this end in every time step the box length
$\lam(t)=\lam_0+t(\lam_1-\lam_0)/\tau$ is changed linearly and all particle
positions are rescaled accordingly. At the end of the compression step, the
system is allowed to relax at the higher density for a time $\tr$ before it is
expanded during the same switching time $\tau$ to reach the initial box size,
after which the cycle is repeated.

\begin{figure}[t]
  \centering
  \includegraphics{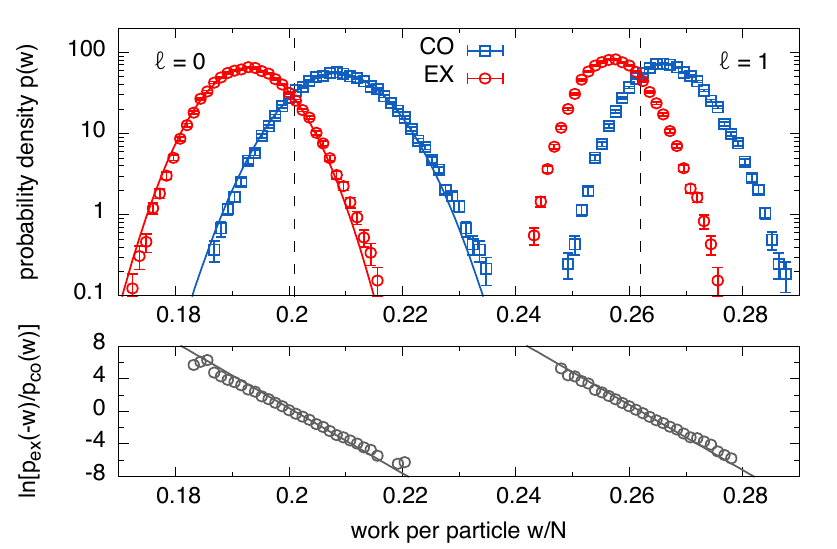}
  \caption{(Top)~Work probability distributions $\pc(w)$ for the compression
    (CO) and $\pe(-w)$ for the expansion (EX) of a two-dimensional model
    suspension between densities $\rho_0=0.1$ and $\rho_1=0.2$. Shown are the
    distributions for the microscopic density (left, $\ell=0$) and
    coarse-grained densities (right, $\ell=1$) using Eq.~\eqref{eq:rho}. The
    solid lines are Gaussian fits to the centers of the $\ell=0$
    distributions. The vertical dashed lines indicate the free energy
    difference $\Delta\Phi_\ell/N$ calculated from
    Eq.~\eqref{eq:bennett}. (Bottom)~Ratio $\ln[\pe(-w)/\pc(w)]$. The solid
    lines show $\Delta\Phi_\ell-w$ for the two values of $\ell$.}
  \label{fig:dist}
\end{figure}

In Fig.~\ref{fig:dist}, the two distributions $\pc(w)$ and $\pe(-w)$ are shown
for 27000 cycles without ($\ell=0$) and with coarse-graining (for $\ell=1$)
for switching time $\tau=0.1$ and equilibration time $\tr=4$. The centers of
the distributions can be fitted well with a Gaussian, whereas the tails
slightly diverge from the Gaussian shape. Under coarse-graining, the work
distributions are shifted and narrower, demonstrating that the coarse-graining
diminishes fluctuations. From the recorded work values for both values of
$\ell$ we calculate the change of free energy $\Delta\Phi$ using
Eq.~\eqref{eq:bennett}.

For the following discussion we make the dependence of the work $w_\ell$ on
the coarse-graining length $\ell$ explicit. The joint probabilities of actual
and coarse-grained work for compression and expansion obey the fluctuation
theorem~\cite{seif12}
\begin{equation}
  \frac{\pe(-w_\ell,-w_0)}{\pc(w_\ell,w_0)} = e^{-w_0+\Delta\Phi}.
\end{equation}
From these joint probabilities, the Crooks work relation~\cite{croo99}
\begin{equation}
  \label{eq:crooks}
  \ln\frac{\pe(-w_0)}{\pc(w_0)} = \Delta\Phi - w_0
\end{equation}
can be derived straightforwardly. In particular, the work value $w_\ast$ where
both distributions cross equals the change of free energy,
$w_\ast=\Delta\Phi$, which is confirmed in Fig.~\ref{fig:dist}.

Assuming that we do not have access to the microscopic work but only to the
coarse-grained work, the fluctuation theorem for the marginal probabilities
becomes
\begin{equation}
  \frac{\pe(-w_\ell)}{\pc(w_\ell)} =
  e^{\Delta\Phi}\IInt{w_0}{-\infty}{+\infty} \pc(w_0|w_\ell) e^{-w_0},
\end{equation}
where $\pc(w_0|w_\ell)$ is the conditional probability to observe an actual
work value $w_0$ given that the coarse-grained work is $w_\ell$. At least for
small $\ell$ the fluctuations of $w_0$ among the trajectories yielding the
same $w_\ell$ can be expected to be sharply peaked around $w_\ell-\mean{\delta
  w_\ell}$ with differential work $\delta w_\ell\equiv w_\ell-w_0$. Assuming
that $\delta w_\ell$ is independent of $w_\ell$, the fluctuation theorem
\begin{equation}
  \label{eq:ft:coarse}
  \ln\frac{\pe(-w_\ell)}{\pc(w_\ell)} = \Delta\Phi + \ln\mean{e^{\delta w_\ell}}
  - w_\ell \equiv \Delta\Phi_\ell - w_\ell
\end{equation}
for the coarse-grained work follows. In contrast to the Crooks relation for
the actual work Eq.~\eqref{eq:crooks}, now the apparent change of the
Helmholtz free energy $\Delta\Phi_\ell$ enters. In Fig.~\ref{fig:dist} it is
shown that Eq.~\eqref{eq:ft:coarse} indeed holds. Note that the straight lines
with slopes $-1$ are not fits but use the value of $\Delta\Phi_\ell$
calculated using Eq.~\eqref{eq:bennett} for the two sets of work values.

\begin{figure}[t]
  \centering
  \includegraphics{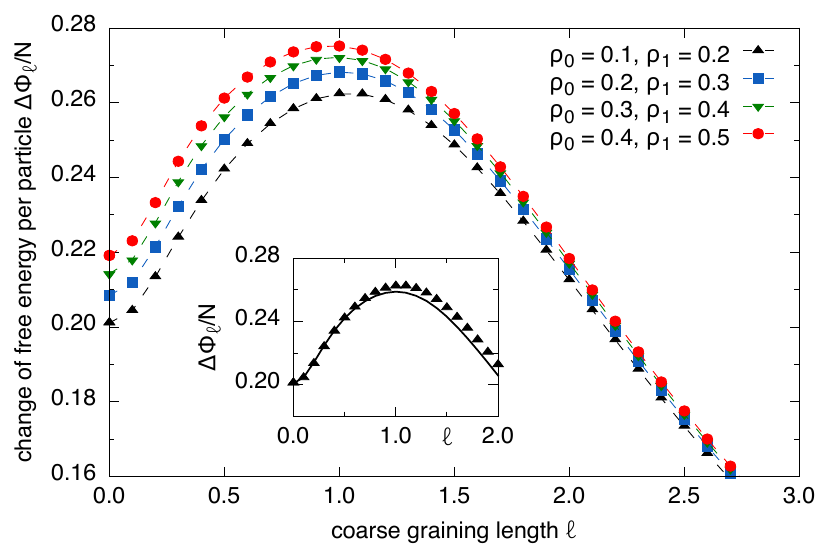}
  \caption{(Apparent) change of free energy per particle $\Delta\Phi_\ell/N$
    as a function of coarse-graining length $\ell$ for several low ($\rho_0$)
    and high ($\rho_1$) densities. The dashed lines are guides to the eye. The
    inset shows the change $\Delta\Phi_\ell/N$ (symbols) together with the
    lower bound $\Delta\Phi/N+\mean{\delta w_\ell}/N$ (solid line).}
  \label{fig:delF}
\end{figure}

In Fig.~\ref{fig:delF} the apparent change of free energy $\Delta\Phi_\ell$ is
plotted as a function of the coarse-graining length $\ell$ for several initial
and final densities. For small $\ell$ the observed free energy increases while
for large coarse-graining length it decreases again and even drops below the
actual value $\Delta\Phi$. Employing the Jensen inequality
$\mean{e^x}\geqslant e^{\mean{x}}$, we obtain the lower bound
\begin{equation}
  \Delta\Phi_\ell \geqslant \Delta\Phi + \mean{\delta w_\ell}.
\end{equation}
As shown in the inset of Fig.~\ref{fig:delF}, for the system studied here this
bound is already a good approximation. We can, therefore, understand the
non-monotonous dependency of $\Delta\Phi_\ell$ from the mean differential work
and, consequently, from the behavior of $f_\ell(r)$ plotted in
Fig.~\ref{fig:press}: For small $\ell$, the effective two-body pressure
$f_\ell(r)>f(r)$ is increased for typical particle distances, leading to
larger work values. Increasing $\ell$ to be larger than the typical particle
distance, the values of $f_\ell(r)$ sampled are typically smaller than $f(r)$
and consequently also the work values are smaller compared to the actual work.


\section{Conclusions}

To summarize, we have studied the stochastic thermodynamics of density fields
$\rho(\x)$ originating from a coarse-graining procedure of microscopic
particle positions. To this end, we have assumed that the Dean
equation~\eqref{eq:dean}, originally derived for the microscopic density, also
holds for coarse-grained density fields. While there is a one-to-one mapping
between particle positions and microscopic density, this is not necessarily
the case for coarse-grained ``smeared'' density profiles, where information is
lost. A more rigorous route to obtain the evolution equation would be to
employ the Mori-Zwanzig projection formalism~\cite{zwan61,grabert}. The
Kawasaki-Dean equation is a, on this level uncontrolled, Markovian
approximation that is, however, \emph{thermodynamically consistent}. By this
we mean that employing the constrained free energy functional $F[\rho]$ that
determines the weight of equilibrium fluctuations to generate both the
dynamics and the work, the Jarzynski relation Eq.~\eqref{eq:jr} holds and
yields the correct change of the free energy for systems driven by
time-dependent protocols. Note that a more general functional form for the
mobility $D[\rho]$ in Eq.~\eqref{eq:dean} (instead of just $D[\rho]=\rho$)
will not change this result as long as the noise obeys the
fluctuation-dissipation theorem, i.e., the noise correlations are proportional
to $D$. Such general mobilities arise in the \emph{macroscopic fluctuation
  theory} (see Ref.~\citenum{derr07} and references therein), which allows to
calculate the large deviation functional (the ``free energy'') for the density
profile in non-equilibrium steady states.

The functional $F[\rho]$ is not known in general. Although it is in principle
different from the DFT density functional $\mathcal F[\rho]$, both have to be
constructed approximately. Since in the limit of small fluctuations the most
probable density profile minimizing $F[\rho]$ will be close to the equilibrium
profile minimizing $\mathcal F[\rho]$, both functionals will practically be
identical. Our results might thus open a route to also investigate and improve
the thermodynamic consistency of density functionals and to use dynamical
density functional theory not only for relaxing but also for driven dynamics.

As a first step in this direction, we have studied the compression and
expansion of a two-dimensional model colloidal suspension. We construct the
density field Eq.~\eqref{eq:rho} as a sum of Gaussians with width $\ell$
centered at the particle positions. As approximation for the density
functional, we employ the quadratic form Eq.~\eqref{eq:U}. While the
microscopic dynamics is governed by the pair potential $u(r)$, the work is
calculated from an effective pair potential $u_\ell(r)$ that depends on the
coarse-graining length. We employ the Yukawa potential, for which the integral
Eq.~\eqref{eq:u} can be performed. For potentials with a steeper repulsion, a
microscopic cut-off has to be employed. The free energy difference extracted
using Eq.~\eqref{eq:bennett} is a non-monotonous function of the
coarse-graining length and can be rationalized from the functional form of the
effective potential. Moreover, we have shown that quite general insights into
the work distributions of coarse-grained processes can be obtained from joint
probabilities of both microscopic and coarse-grained work. In particular, the
Jarzynski relation and the Crooks relation Eq.~\eqref{eq:ft:coarse} hold
involving the change of an apparent free energy that depends on the
coarse-graining length. This demonstrates that care has to be taken: Even if
the fluctuation theorem exhibits the correct slope, the free energy change
might be systematically effected by measurement uncertainties.


\appendix

\section{Antisymmetric part of the stochastic action}
\label{sec:ft}

The derivation presented here follows standard arguments for Gaussian noise
(see, e.g., Ref.~\cite{schm07a}) starting with Eq.~\eqref{eq:dean}. In the
following, it will become convenient to define the scalar noise
$\zeta(\x,t)\equiv\nabla\cdot\nois(\x,t)$ with correlations
\begin{equation}
  \label{eq:K}
  \begin{split}
    K(\x,t|\x',t') &\equiv \mean{\zeta(\x,t)\zeta(\x',t')} \\
    &= 2T\delta(t-t')\nabla\cdot\nabla'\rho(\x,t)\delta(\x-\x'),
  \end{split}
\end{equation}
where $\nabla'$ acts on $\x'$. The probability of a noise history is Gaussian,
$\mathcal P[\zeta]=e^{-\mathcal A[\zeta]}$, with stochastic action
\begin{equation}
  \label{eq:A}
  \mathcal A[\zeta] \equiv \frac{1}{2}\Int{\x\dd t\dd\x'\dd t'}
  \zeta(\x,t)K^{-1}(\x,t|\x',t')\zeta(\x',t').
\end{equation}
The kernel $K^{-1}$ is the inverse of the noise correlations Eq.~\eqref{eq:K}
in the operator sense,
\begin{multline}
  \Int{\x''\dd t''} K(\x,t|\x'',t'')K^{-1}(\x'',t''|\x',t') \\
  = \delta(t-t')\delta(\x-\x').
\end{multline}
Inserting Eq.~\eqref{eq:K} and integrating by parts, it is easy to see that
the gradient of the inverse kernel can be written as
\begin{equation}
  \label{eq:K:inv}
  \nabla K^{-1}(\x,t|\x',t') = \frac{1}{2T}
  \frac{\delta(t-t')}{\rho(\x,t)}\nabla G(\x-\x'),
\end{equation}
where $G(\x-\x')$ is the Green's function of the Laplace operator,
\begin{equation}
  \label{eq:G}
  \nabla^2 G(\x-\x') = -\delta(\x-\x').
\end{equation}

We now rearrange the evolution equation~\eqref{eq:dean} for the density,
\begin{equation}
  \zeta(\x,t) = \partial_t\rho(\x,t) -
  \nabla\cdot\rho(\x,t)\nabla\fd{F[\rho]}{\rho(\x,t)}.
\end{equation}
We insert this expression into Eq.~\eqref{eq:A} to obtain the stochastic
action as a function of density histories. Note that changing the fields
$\zeta\mapsto\rho$ implies a Jacobian. However, we are only interested in the
part of the action that is antisymmetric with respect to time reversal, which
reads
\begin{widetext}
  \begin{equation}
    \begin{split}
    \mathcal A[\Gam^\dagger;\lam^\dagger] - \mathcal A[\Gam;\lam] 
    &= 2\Int{\x\dd t\dd\x'\dd t'}
    \left\{\nabla\cdot\rho(\x,t)\nabla\fd{F[\rho]}{\rho(\x,t)}\right\}
    K^{-1}(\x,t|\x',t')\partial_t\rho(\x',t') \\
    &= -2\Int{\x\dd t\dd\x'\dd t'}
    \rho(\x,t)\left[\nabla\fd{F[\rho]}{\rho(\x,t)}\right]
    \cdot\nabla K^{-1}(\x,t|\x',t')\partial_t\rho(\x',t') \\
    &= -\frac{1}{T}\Int{\x\dd t}\fd{F[\rho]}{\rho(\x,t)}\partial_t\rho(\x,t)
    = -\frac{q_0}{T},
    \end{split}
  \end{equation}
\end{widetext}
where we have used Eqs.~\eqref{eq:K:inv} and~\eqref{eq:G}. This is the first
term in Eq.~\eqref{eq:R:res}. The full weight of a trajectory reads $\mathcal
P[\Gam;\lam]=\psi_0[\rho(\x,0)]e^{-\mathcal A[\Gam;\lam]}$, leading to the
second term in Eq.~\eqref{eq:R} as boundary term for trajectories starting
with density profile $\rho(\x,0)$ (forward protocol) and $\rho(\x,\tau)$
(backward protocol). The antisymmetric part of the action is thus related to
the heat dissipated into the environment as expected.


\end{document}